\documentstyle[psfig]{mn}

\newif\ifAMStwofonts

\def\kms{\thinspace\hbox{$\hbox{km}\thinspace\hbox{s}^{-1}$}}
\def\Mpc{\thinspace\hbox{Mpc}}

\def\kms{\thinspace\hbox{$\hbox{km}\thinspace\hbox{s}^{-1}$}}
\def\hb{\hbox{$\hbox{H}\beta$}}
\def\lbeta{\thinspace\hbox{L(H$\beta$)}}
\def\fbeta{\thinspace\hbox{F(H$\beta$)}}
\def\abeta{\thinspace\hbox{A(H$\beta$)}}

\def\hbeta{\thinspace\hbox{H$\beta$}}
\def\hgamma{\thinspace\hbox{H$\gamma$}}
\def\wbeta{\thinspace\hbox{W(H$\beta$)}}

\def\etal{{\it et al.}}      

\font\elevenit=cmti10 scaled 1090
\def\kms{\thinspace\hbox{$\hbox{km}\thinspace\hbox{s}^{-1}$}}
\def\kmsec{\thinspace\hbox{$\hbox{km}\thinspace\hbox{s}^{-1} $}}
\def\hnot{\thinspace\hbox{H$_{\mbox{\scriptsize o}}$}}
\def\hunits{\thinspace\kmsec\Mpc$^{-1}$}

\ifoldfss
  \ifCUPmtlplainloaded \else
    \NewTextAlphabet{textbfit} {cmbxti10} {}
    \NewTextAlphabet{textbfss} {cmssbx10} {}
    \NewMathAlphabet{mathbfit} {cmbxti10} {} 
    \NewMathAlphabet{mathbfss} {cmssbx10} {} 
  \fi
  \ifAMStwofonts
    \ifCUPmtlplainloaded \else
      \NewSymbolFont{upmath} {eurm10}
      \NewSymbolFont{AMSa} {msam10}
      \NewMathSymbol{\upi}     {0}{upmath}{19}
      \NewMathSymbol{\umu}     {0}{upmath}{16}
      \NewMathSymbol{\upartial}{0}{upmath}{40}
      \NewMathSymbol{\leqslant}{3}{AMSa}{36}
      \NewMathSymbol{\geqslant}{3}{AMSa}{3E}

      \let\leq=\leqslant 
       
    \fi
  \fi
\fi 

\ifnfssone
  \newmathalphabet{\mathit}
  \addtoversion{normal}{\mathit}{cmr}{m}{it}
  \addtoversion{bold}{\mathit}{cmr}{bx}{it}
  \newmathalphabet{\mathbfit} 
  \addtoversion{normal}{\mathbfit}{cmr}{bx}{it}
  \addtoversion{bold}{\mathbfit}{cmr}{bx}{it}
  \newmathalphabet{\mathbfss} 
  \addtoversion{normal}{\mathbfss}{cmss}{bx}{n}
  \addtoversion{bold}{\mathbfss}{cmss}{bx}{n}
  \ifAMStwofonts
    \ifCUPmtlplainloaded \else
      \UseAMStwoboldmath
      \makeatletter
      \new@mathgroup\upmath@group
      \define@mathgroup\mv@normal\upmath@group{eur}{m}{n}
      \define@mathgroup\mv@bold\upmath@group{eur}{b}{n}
      \edef\UPM{\hexnumber\upmath@group}
      \new@mathgroup\amsa@group
      \define@mathgroup\mv@normal\amsa@group{msa}{m}{n}
      \define@mathgroup\mv@bold\amsa@group{msa}{m}{n}
      \edef\AMSa{\hexnumber\amsa@group}
      \makeatother
      \mathchardef\upi="0\UPM19
      \mathchardef\umu="0\UPM16
      \mathchardef\upartial="0\UPM40
      \mathchardef\leqslant="3\AMSa36
      \mathchardef\geqslant="3\AMSa3E

      \let\leq=\leqslant 

    \fi
  \fi
\fi 

\ifnfsstwo
  \DeclareMathAlphabet{\mathbfit}{OT1}{cmr}{bx}{it}
  \SetMathAlphabet\mathbfit{bold}{OT1}{cmr}{bx}{it}
  \DeclareMathAlphabet{\mathbfss}{OT1}{cmss}{bx}{n}
  \SetMathAlphabet\mathbfss{bold}{OT1}{cmss}{bx}{n}
  \ifAMStwofonts
    \ifCUPmtlplainloaded \else
      \DeclareSymbolFont{UPM}{U}{eur}{m}{n}
      \SetSymbolFont{UPM}{bold}{U}{eur}{b}{n}
      \DeclareSymbolFont{AMSa}{U}{msa}{m}{n}
      \DeclareMathSymbol{\upi}{0}{UPM}{"19}
      \DeclareMathSymbol{\umu}{0}{UPM}{"16}
      \DeclareMathSymbol{\upartial}{0}{UPM}{"40}
      \DeclareMathSymbol{\leqslant}{3}{AMSa}{"36}
      \DeclareMathSymbol{\geqslant}{3}{AMSa}{"3E}

      \let\leq=\leqslant 

    \fi
  \fi
\fi 

\ifCUPmtlplainloaded \else
  \ifAMStwofonts \else 
    \def\upi{\pi}
    \def\umu{\mu}
    \def\upartial{\partial}
  \fi
\fi
\title[HII Galaxies as deep cosmological probes]{HII Galaxies as deep
cosmological probes}
\author[Jorge Melnick et al.]
       {Jorge Melnick$^{1}$ \thanks{jmelnick@eso.org}, 
Roberto Terlevich$^2$ 
\thanks{Visiting Professor, INAOE, Puebla,
Mexico}  
and Elena Terlevich $^3$ \thanks{Visiting Fellow, IoA, Cambridge}
\cr 
 \\
        $^{1}$ European Southern Observatory, Alonso de Cordova 3107, 
Santiago, Chile.\\
        $^{2}$  Institute of Astronomy, Madingley Road, Cambridge CB3 0HA, 
UK.\\
        $^{3}$ Instituto Nacional de Astrof\'\i sica, Optica y
Electr\'onica,
A.P. 51 y 216,  72000 Puebla, Mexico \\
       }

\date{Accepted ....
      Received ...;
      in original form ...}
\pubyear{1999}
\begin{document}
\include{macro.tex}
\maketitle

\begin{abstract}

We re-investigated the use of the Hubble diagram to measure the cosmological 
constant ($\Lambda$)  and the mass density  of the Universe ($\Omega_M$). 
We find an important focusing effect in $\Lambda$ for redshifts about 3. This 
effect implies that the apparent magnitude of a standard candle at redshifts 
z=2-3 has almost no dependence on $\Lambda$ for   $\Omega_M>0.2$.
This means that $\Omega_M$ can be measured independently of
$\Omega_{\Lambda}$ by targeting the redshift range according to 
an estimate of the value of $\Omega_M$.

We explore the evidence in support of the suggestion that extreme starburst 
galaxies also known as HII galaxies can be used as distance estimators over 
a wide range of redshifts and reaching very high values. We have compiled 
literature data of HII galaxies up to $z\sim 3$ and found a good correlation 
between their luminosity and velocity dispersion measured from their strong 
emission lines, thus confirming the correlation already known to exist for 
HII galaxies in the nearby Universe. Several systematic effects such as age, 
extinction, kinematics, and metallicity 
are discussed as well as the effects of different cosmologies.

\end{abstract}


\section{Introduction}

Recent results from distant supernova surveys have yielded
values of $\Omega_M$ (the matter density parameter of the Universe) 
so low (in fact negative for $\Lambda = 0$) that they seem 
unphysical and in disagreement
with CMB results. This inconsistency has lead to a renewed
exploration of cosmological models with cosmological constant  
$\Lambda$ (Lineweaver, 1998; White, 1998).

Non-zero $\Lambda$ has been invoked before to solve inconsistencies
or apparent discrepancies, such as  the expansion age
of the Universe versus the age of globular clusters, but 
the most compelling evidence for $\Lambda\neq0$ comes from 
the combination of the observed CMB anisotropy and the constraints from 
distant type Ia supernovae
(see Efstathiou \etal, 1999 for a recent review).

The use of supernovae to  measure simultaneously 
$\Omega_{M}$ and $\Omega_{\Lambda}$ (the energy density of vacuum) 
was pioneered by Goobar and Perlmutter (1995) and nicely 
demonstrated by Perlmutter et al. (1998) and Riess et al. (1998)  who
showed that type Ia SN at redshifts $0.1<z<1$ could strongly constrain the 
allowed range in these cosmological parameters. Unfortunately, the results of 
the two groups are still inconsistent at the $2\sigma$ level, although when
combined 
they tend to favor models with low matter density ($\Omega_{M}\leq0.4$) and
non-zero $\Lambda$ (Efstathiou and Bond 1999; Efstathiou \etal, 1999).

In this paper we show that the strong {\it focusing effect} of Hubble 
diagrams with cosmological constant allows to separate cleanly the effects of 
mass density and vacuum density in the expansion, provided one can measure 
distances in the range $1<z<3$. At $z=2-3$ the discrimination between 
different values of $\Omega_{M}$ reaches up to one magnitude in distance 
modulus, and is only very 
weakly dependent on $\Omega_{\Lambda}$, while \it knowing $\Omega_{M}$, \rm 
the discrimination in $\Omega_{\Lambda}$ is largest for $z=0.6-1$ and reaches 
about 0.5~mag at $z\sim 1$ for $\Omega_{M}=0.2$.

It is therefore desirable to explore distance estimators like the 
$\rm \lbeta-\sigma$ relation in HII galaxies (Melnick et al 1988, hereafter 
MTM) that can  be potentially used  from the local group of galaxies up to 
z~$\sim 4$ with today's technology.  In this {\em Letter} we
use published data to show that the $\rm \lbeta-\sigma$ relation
for local galaxies is also satisfied by emission line objects of
redshifts up to $z \simeq 3$.  We argue  that strong emission line
galaxies are very promising objects to be used for a global determination of 
the cosmological parameters $\Omega_{M}$ and $\Omega_{\Lambda}$.

\section[]{The redshift-magnitude diagram in $\Lambda \ne 0$ cosmologies}

The emission line luminosity of an object is related to the 
observed emission line flux through 
the luminosity distance parameter $D_L$ that depends on the cosmological
parameters $\Omega_{M}$ and $\Omega_{\Lambda} = \Lambda / (3H^2_0)$ as
(Refsdal \etal\ 1967):

\vskip 2.mm
\noindent
$D_L=\frac{c(1+z)}{H_0|\Omega_{\kappa}|^{1/2}}{\elevenit F {\elevenit [} |\Omega_{\kappa}|^{1/2} 
I(z,\Omega_{M}, \Omega_{\Lambda}){\elevenit ]}  } $
\vskip 2.mm

\noindent
$ I (z,\Omega_{M}, \Omega_{\Lambda}) = \int _0 ^z [ (1+z')^2(1+\Omega_{M} z')- 
z'(2+z')\Omega_{\Lambda}]^{-1/2}dz' $

\vskip 2.mm

\noindent
where 
\vskip 2.mm

\noindent
{ F[x]} is sin(x) for $\Omega_{M} + \Omega_{\Lambda} < 1$
\vskip 2.mm

\noindent
{F[x]} is sinh(x) for $\Omega_{M} + \Omega_{\Lambda} > 1$
\vskip 2.mm

\noindent
and $\Omega_{\kappa}=1-\Omega_{M}-\Omega_{\Lambda}$ in both cases.
\vskip 2.mm

\noindent
{ F[x]}$= x$ and $\Omega_{\kappa}=1$ for $\Omega_{M} + \Omega_{\Lambda} = 1$
\vskip 2.mm

For $D_L$ in Mpc, the relation between apparent (m) and absolute (M) 
emission-line or continuum magnitudes is given by,
\vskip 2.mm

\noindent
$m = M + 5 log D_L + 25$

\vskip 3.mm

\begin{figure}
\psfig{figure=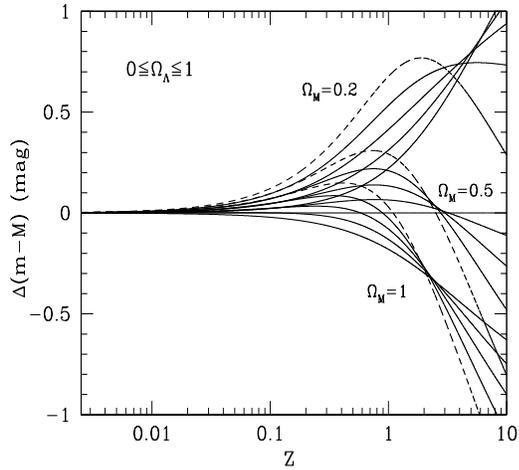,height=8cm,width=7cm,angle=0}
\caption
{Normalized distance modulus $\Delta(m-M) = (m-M)_{\Omega_{M},\Omega_{\Lambda}} - (m-M)_{.5,0}$
as a function of redshift. For each value of 
$\Omega_{M}$ as labeled in the Figure we plot a family of vacuum energy density
$\Omega_{\Lambda}=0,0.25,0.5,0.75,1.0$. For each family, the dashed line 
corresponds to $\Omega_{\Lambda}=1$.}
\label{focus} 
\end{figure}

An important and perhaps surprising feature of the Hubble diagrams with 
non-zero cosmological constant is the strong \it focusing or convergence 
effect \rm mentioned by Refsdal \etal\ (1967).  This is shown in 
Figure~\ref{focus} that plots the predicted luminosity distance (normalized 
to $\Omega_{M}=0.5$ and $\Omega_{\Lambda}=0$) as a function of redshift for 
different combinations of cosmological parameters.   For a given $\Omega_{M}$ 
the world models of different $\Omega_{\Lambda}$ converge in a narrow redshift 
range and the degree of convergence increases with increasing mass density.  
The redshift at which the convergence occurs diminishes with increasing values 
of $\Omega_{M}$. In particular, for $\Omega_{M}=0.5$
the convergence redshift is about 2.8; for $\Omega_{M}=1.0$ it is about 2.3
and for $\Omega_{M}=2.0$, about 1.7. For $ \Omega_{M} = 0$ (not shown in the 
figure) there is no convergence while for $\Omega_{M}<0.2$  the critical 
redshift is $z>>10$.

As discussed in the Introduction, this effect allows the accurate 
determination of $\Omega_{M}$
independently of the value of $\Omega_{\Lambda}$. For small $\Omega_{M}$
the optimum redshift range is $z\sim 3$ where already a large sample of
HII galaxies does exist. The existence of this focusing also implies that
the best range to determine  $\Omega_{\Lambda}$ using the magnitude-redshift
method is either $z<1$ or $z>5$ .

\section []{The distance estimators with wide redshift range}

The classical empirical distance estimators for spiral and elliptical
galaxies (Tully-Fisher and $\rm D_n-\sigma$) cannot be applied to galaxies
at large redshifts (say $z>0.5$) because of significant systematic
evolution of the stellar populations with look-back time (Schade {\etal\ }
1997, Rix \etal\  1997, Vogt \etal\  1997, Van Dokkum \etal\ 1998).  Thus,
even if we could measure the relevant parameters with the next
generation of ground-based and space telescopes, it is still unclear whether
it will be possible to use these techniques to determine reliable distances
of galaxies at $z>0.5$.

Supernovae Ia are good standard candles (see e.g.~Perlmutter \etal\ 1998; 
Riess \etal\ 1998, and references therein) with errors less than 0.4 mag for 
a single SN up to  redshifts of about 0.8. There are, however, discrepancies 
between the results of these two groups which may be related to uncertainties 
in the extinction corrections or other as yet unknown systematic effects such 
as metallicity. Nevertheless, SNIa is still the most accurate method 
that can be used up to z $\sim 1$ with present day instrumentation.

A potentially very powerful technique that has received relatively little 
attention in the literature is the L($H_{\beta}$)-$\sigma$ relation for HII 
galaxies.  The correlation between the $H_{\beta}$ luminosity (\lbeta) and 
the velocity width of the lines ($\sigma$) described by Terlevich and 
Melnick (1981) was calibrated as a distance indicator on a sample of nearby 
galaxies ($z<0.1$) by Melnick \etal, (1987, MTM) using giant 
HII regions in nearby late type galaxies to fix the zero point. Since the 
(bolometric) luminosities of HII galaxies 
are dominated by one or more starburst components, their luminosities per unit 
mass are very large. So, in spite of being low mass objects, HII galaxies can 
be observed out to redshifts of cosmological interest.

By selecting star-forming galaxies with the strongest emission lines 
(i.e.~ with the largest equivalent widths), 
one effectively selects the youngest objects within a 
narrow age range (Copetti, Pastoriza \& Dottori 1986). This selection 
criterion guarantees that, at least to first order, the $\rm \lbeta-\sigma$ 
distance estimator is free from the evolution effects in the
stellar population which bedevil the traditional techniques. Moreover, the 
extinction and the metallicity of these galaxies can be directly determined 
from their emission line spectra. So even possible systematic changes
in metallicity with redshift, for example, can be included in a relatively 
straightforward way since oxygen abundances can be directly determined 
with the new generation of IR spectrographs on 8m class telescopes.
Although HII galaxies are to first order free from
the systematic effects that plague SNIa, the 
error  in the distance modulus for a given galaxy in the current 
calibration is about
twice that of SNIa.  However, most of the scatter in the correlation is due 
to observational
errors which can be substantially reduced using modern instruments and 
detectors.

The $\lbeta-\sigma$ relation has recently been verified to hold also for star 
forming faint blue galaxies at redshifts of about $z\sim0.5$ (Koo \etal\  
1996) which
constitutes a first very important step towards its use as a deep cosmological 
probe.

\section[]{The HII galaxies dataset}

The sample used by  MTM to calibrate the distance indicator is limited to 
$z<0.1$.  In order to extend this sample to distances of cosmological
interest we have searched the literature for galaxies at $z>0.1$ having 
very strong and narrow emission lines. Unfortunately, such
objects are rare in catalogues of faint blue galaxies at intermediate
redshifts, 
but many objects with moderate emission line strengths have been found in
deep photometric searches.  Koo \etal (1994, 1995) and Guzm\'an \etal 
(1996, 1998) have published
images and high resolution spectra of 17 faint (B=20--23)
blue galaxies with narrow lines at redshifts
between 0.1 and 1.  Their images, spectra, luminosities, and line
widths are a close match to those of the nearby
HII galaxies, so Koo and collaborators concluded that HII galaxies
are the local counterparts of their intermediate redshift
compact blue galaxies.

Guzm\'an \etal\  (1997) published data for 51 compact galaxies in 
the Hubble deep field (HDF) of which 27 are classified as ``HII-like''. They 
(and also Koo \etal ) give emission-line widths ($\sigma$), absolute
blue magnitudes ($\rm M_B$), and \hb\  equivalent widths (\wbeta)
obtained with the Keck~I telescope.  The authors also give \hb\
luminosities which they  derive from the absolute blue magnitudes
and equivalent widths following Terlevich and Melnick (1981).

A few Lyman-break galaxies show strong emission lines.  Pettini \etal\  
(1998) have presented near IR spectroscopy of 5 Lyman break
galaxies at $z\sim3$.  Two of these galaxies have luminosities
and velocity dispersions typical of HII galaxies. A third one has
very strong lines ($\wbeta>50$\AA) but the velocity dispersion
($\sigma=190\kms$) is too large for HII galaxies (see below).
No \hb\  fluxes or equivalent widths are available for the remaining
two objects.

\begin{figure}
\psfig{figure=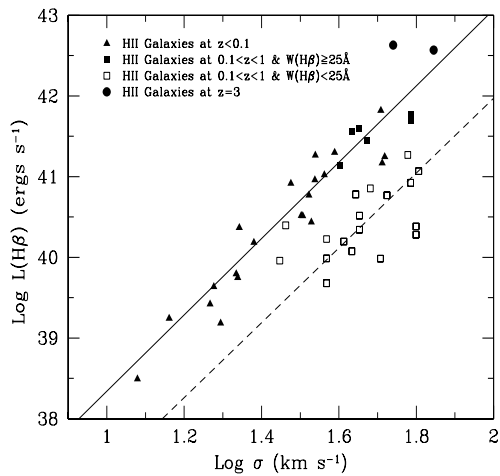,height=8cm,width=7cm,angle=0}
\caption
{The Luminosity-sigma correlation for HII galaxies at a wide
range of redshifts.  The solid line shows the Maximum-Likelihood fit
to the young HII galaxies in the local Universe. The dashed line
shows the predicted $\lbeta-\sigma$ relation for an evolved population
of HII galaxies. The cosmology is $\rm H_o=65, q_o=0, \Lambda=0$ in
this Figure. } 
\label{lsigma} 
\end{figure}

Figure~\ref{lsigma} shows the $\lbeta-\sigma$ relation for the
galaxies in the samples above as follows: filled triangles present the data 
for local galaxies from MTM.  Squares show the data from Koo \etal, (1995) 
and Guzm\'an \etal, (1997). The circles present the high redshift objects from 
Pettini \etal, (1998).  The solid line shows the MTM fit to the local objects.
Because no extinction measurements are available for the intermediate and  high
redshift samples, the local sample galaxies are plotted in Figure~\ref{lsigma} 
without extinction corrections.  

\section[]{The $\lbeta-\sigma$ relation as a distance indicator}

The most important systematic effects in the $\lbeta-\sigma$ relation that 
need to be considered in order to apply the correlation 
to high redshift galaxies are summarized below.

\subsection{The physics of the $\lbeta-\sigma$ relation}

There has been considerable debate in the literature concerning the 
interpretation of the emission line-profile widths in Giant HII Regions 
(GHR) which in many respects resemble HII galaxies and which, in particular, 
exhibit a similar correlation between $\lbeta$ and $\sigma$.  In GHR, the 
coupling between the turbulence of the ionized gas ($\sigma$), and the total 
mass of the system (stars + gas) is very complex and appears to evolve with 
time.  So for young GHR $\sigma$ is coupled to gravity through the stirring 
motions of low-mass stars, while for evolved objects this coupling is lost
and the gas motions are dominated by stellar winds from massive stars 
(see Melnick, Tenorio-Tagle, and Terlevich, 1999 for a recent review).  It is 
still not known if age is the only (or the dominant) parameter, or if 
environment also plays an important
role, but the fact that GHR with a wide range of ages
fit the $\lbeta-\sigma$ relation suggests
that the total mass of the objects is what determines $\sigma$.

The situation for HII galaxies is different.  Telles (1995) showed that 
these objects define a fundamental plane that is remarkably similar to that 
defined by normal elliptical galaxies.  This result lends strong support to 
the interpretation of Terlevich and Melnick (1981) and MTM that the emission 
line-profile widths of Giant HII galaxies directly measure the total mass of 
these systems within the measuring radius.  Therefore, besides systematic 
effects, that are discussed below, the scatter in the $\lbeta-\sigma$ 
(notice that Telles used continuum magnitudes and not $\lbeta$) depends among 
other things on the existence of a second parameter (see below), on possible 
variations of the IMF, on the importance of sources of broadening not related 
to a young stellar component (e.g. rotation), and on
the duration of the burst of star-formation that powers the emission lines.

\begin{figure}
\psfig{figure=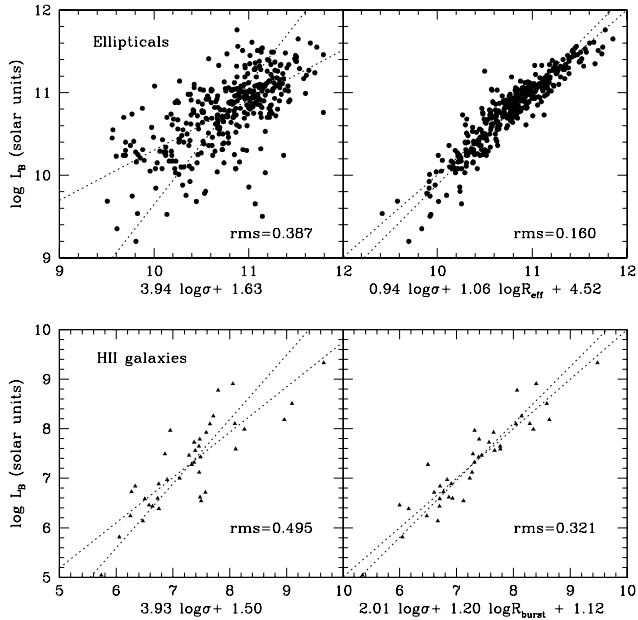,height=9cm,width=9cm,angle=0}
\caption
{The fundamental-plane of HII galaxies and normal elliptical galaxies from
Telles (1995).
The radii and magnitudes of HII galaxies are measured from continuum
images. The velocity dispersions are the widths of the emission lines. }
\label{edu} 
\end{figure}

MTM showed that this scatter can be reduced by restricting the sample to 
objects with $\sigma<65$\kms.
The same result was found by Koo \etal,  (1995) for intermediate redshift
objects.  This cutoff can be
understood if one assumes that HII galaxies are powered by clusters
of coeval stars (starbursts).  The cutoff results from imposing the condition
that the time required for the clusters to form (e.g. the free-fall time)
must be smaller than the main-sequence life-time of the most massive stars.
One of the two galaxies at z=3 plotted in Figure~\ref{lsigma} appears
to exceed this limit, but the measurement error ($\pm 20$\kms)
is still rather large.

\subsection{Age effects}

In order to minimize systematic effects due to the rapid evolution of the
ionizing stars, MTM  restricted their sample to galaxies with
$\wbeta>25$\AA.  In fact, this restriction has a double purpose which
is particularly relevant for high-z objects: it selects the young(est)
starbursts, and eliminates objects with significant underlying
old(er) stellar populations.  The latter is critical because
an old stellar population may widen the emission lines in a way that is 
uncorrelated with the luminosity of the young component.

There are only a few objects in our intermediate redshift sample
with $\wbeta>25$\AA.  These are plotted with filled symbols in
Figure~\ref{lsigma}.  The open symbols show the data for objects
with weaker lines.  As expected, these objects do not fit the
correlation  defined by the local HII galaxies which have a
mean line strength of $<\wbeta>=105$\AA. 

The luminosity evolution of a young coeval starburst during the
first $10^7$yr proceeds as a rapid decay of the emission line flux
after the first 3~Myr at roughly constant continuum flux until about 6~Myr. 
Thus, in this range of ages the age-dimming in $\lbeta$ can be directly 
estimated from the change in equivalent widths (Terlevich \& Melnick, 1981;
Copetti, Pastoriza \& Dottori, 1986). The mean equivalent width of the objects
plotted as open squares in Figure~\ref{lsigma} is $<\wbeta>=11$\AA\ so 
evolution reduces the average \hb\ luminosity of the sample by a factor 105/11.
The dashed line shows the MTM relation affected by this amount of evolution.
The fit is seen to be more than acceptable, confirming our conclusion that
most objects are evolved starbursts rather than strong starbursts
on top of a bright, older stellar population.  

Note, however, that two of the weak-lined galaxies fit the correlation without
luminosity corrections. These objects have high \hbeta\ luminosities but also 
very strong continua, indicating the presence of a significant underlying 
older stellar population.  The HST images of one of these galaxies (H1-3618) 
by Koo \etal,  (1994) show that this object is very compact indicating that 
the strong continuum does not come from a bright underlying galaxy, but is 
most likely the light from a previous starburst.

Another indication that the objects in the intermediate redshift sample  are in
general more evolved than the local sample ones comes from the excitation
of the nebular gas as measured by the ratio of [OIII]/\hb.  The mean
value from Guzm\'an \etal,  (1997) is $<[OIII]/\hb>=2.2\pm0.5$
while the mean for the MTM sample is $<[OIII]/\hb>=5\pm2$ where the quoted
errors are the ($1\sigma$) widths of the distributions.  

\subsection{Extinction effects}

\begin{figure}
\centerline{\psfig{figure=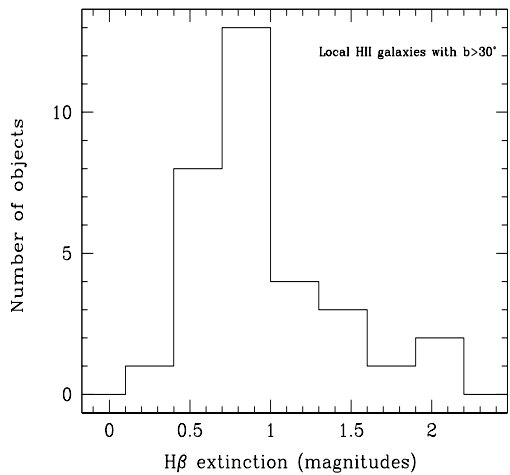,height=8cm,width=7cm,angle=0}}
\caption{Distribution of extinction determined from the Balmer
decrements for local HII galaxies at high galactic latitudes
($b>30^{\circ}$).} \label{extin} \end{figure}

Extinction corrections for local HII galaxies are determined
in a straightforward manner from the Balmer decrements (MTM).
Figure~\ref{extin} presents a histogram of the extinction for
the MTM galaxies at high galactic
latitudes ($b>30^{\circ}$) in order to minimize the contribution
of foreground galactic extinction.  The extinction is strongly
peaked at a value of $A_{\hb}=0.8^m$ with a mean value of $A_{\hb}=1.1^m$
and an rms value of $0.5^m$. Restricting the sample to the luminosity range
covered by intermediate redshift objects (log($\lbeta>41.0$) gives a slightly
larger value $A_{\hb}=1.25^m$ with similar dispersion.

It is rather difficult to measure the Balmer decrement for low
S/N observations of intermediate and high redshift HII galaxies and
it is normally not done, so no extinction values are available for the
intermediate and high redshift galaxies in our sample. Future observations 
with 8m-class telescopes should ideally include $\hbeta$ and 
$\hgamma$ to permit direct estimates of the reddening in high-z HII galaxies.

\subsection{Metallicity effects}

In their calibration of the $\lbeta-\sigma$ relation as
a distance indicator, MTM found an important systematic shift in
luminosity between the giant  HII regions in nearby late type galaxies,
used to determine the zero point, and HII galaxies, which is 
due to differences in the mean metallicities of the two samples.

\begin{figure}
\centerline{\psfig{figure=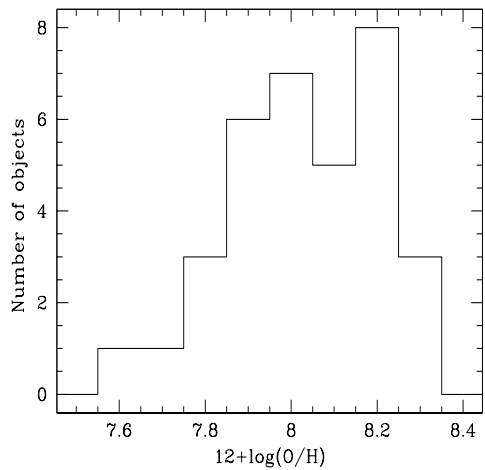,height=8cm,width=7cm,angle=0}}
\caption{Distribution of Oxygen abundances for HII galaxies in the
local Universe ($z<0.1$).} 
\label{abun} 
\end{figure}

The distribution of  metallicities for the MTM HII galaxies in
our sample  is presented in Figure~\ref{abun}.  The mean metallicity of
the sample is $12+log(O/H)=8.02\pm0.18(\sigma)$ while if we restrict the sample
to the most luminous objects as described above the mean metallicity is 
$12+log(O/H)=8.07\pm0.19(\sigma)$. Unfortunately,
there are no metallicities available yet for the intermediate
and high redshift objects in our sample, but clearly, in order to use the
$\lbeta-\sigma$ relation as a distance indicator, either the
metallicities of the local and high-z samples must be similar,
or the luminosities must be corrected using O/H as prescribed by MTM.

In a pilot project to measure accurate metallicities and electron 
temperatures of HII galaxies at redshifts between 0.2 and 1 
(Terlevich \etal, in preparation) we observed a sample of 20 low 
metallicity candidates with the 3.6m and NTT telescopes at La Silla. 
We could clearly detect the  electron temperature sensitive faint line 
[OIII]$\lambda 4363$\AA\ in the spectra of the five objects with the largest 
$\wbeta$. A preliminary analysis of the data yields a mean oxygen abundance 
of $<12+log(O/H)>=7.8$, significantly lower, in fact, than the mean value for 
the local sample.  Recently, Kobulnicky and Zaritsky (1999; KZ99) have
presented data for HIIG-like objects with redshifts $0.1<z<0.3$.  Their mean
abundance $\rm - <12+log(O/H)>=8.4 -$ is significantly larger than what we get
for our NTT sample.  However, KZ99 detect [OIII]$\lambda 4363$\AA\ (and hence
measure electron temperatures) in only two of their objects, but for 
one of them the detection is marginal. The abundances of the other objects, 
estimated using the empirical R$_{23}$ method, cannot be used with any 
confidence because the zero-point offset can be as large as 0.5~dex. The 
abundance of the only object with a good measurement of electron temperature
is 12+log(O/H)=7.84.  Although KZ99 conclude tentatively that O/H in their 
intermediate redshift sample is larger that in the nearby HII galaxies, we 
think that  direct measurements of  electron temperatures are needed to 
support such claim. Empirical methods are based on the underlying assumption 
that the ionizing properties of the young stellar populations are the same in 
the different objects, assumption that must be checked when comparing
objects over a wide range in redshifts.

We conclude that there is tentative, albeit contradictory, evidence that the 
abundances of higher redshift objects could be different from those of local 
HII galaxies.  Although the data are still very sparse and inaccurate, if 
real, such effect would introduce an important systematic bias in the 
estimation of distances to high redshift objects that must be taken into 
account.

\section[]{HII galaxies as cosmological probes}

In order to illustrate the potential of HII galaxies as deep cosmological 
probes we have calculated the predicted distance moduli for the objects
plotted in Figure~\ref{lsigma} using the most recent data from the
literature (distances and oxygen abundances) for the giant HII regions
in order to re-calibrate the zero-point. The new calibration 
of the unbiased distance indicator introduced by MTM, 
$M_Z ={{\sigma^5}\over{O/H}}$, is thus given by, 

$$ log(\lbeta) = log(M_Z) + 29.5 $$ 

\noindent
from which the distance modulus is obtained as:

$$(m-M) = 2.5\times log{{\sigma^5}\over{\fbeta}} - 
2.5\times log(O/H) - \\
A_{\hb} - 26.44 \ \ (1) $$

\noindent
where \fbeta\ is the observed \hbeta\  flux and  
$A_{\hb}$ is the total extinction.  

\begin{figure}
\psfig{figure=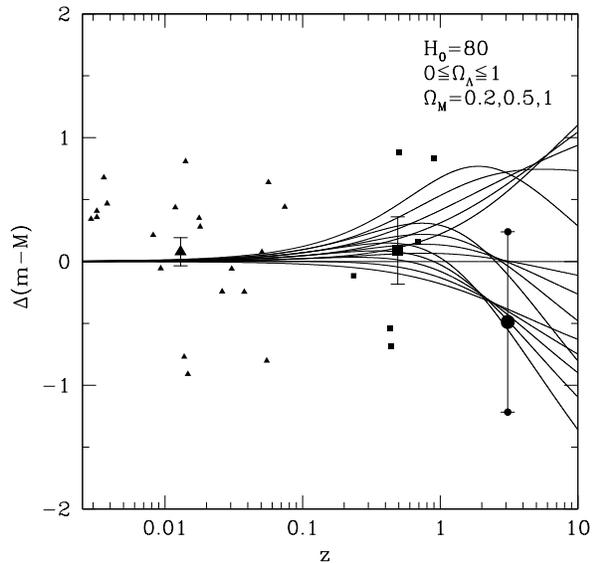,height=8cm,width=8cm,angle=0}
\caption
{The differential Hubble diagram for HII galaxies of a wide range of redshifts. 
The family of curves from Figure~\ref{focus} is also shown. 
The large symbols represent the
average redshift and distance modulus for each sub-sample.  The error-bars 
show the mean error in distance modulus assuming that each data-point is an 
independent measurement and ignoring observational errors. \hnot=80 \hunits was
used to normalize the data points.  The model lines are independent of \hnot.}
\label{hubble} 
\end{figure}

Figure~\ref{hubble} presents the  resulting Hubble diagram for HII galaxies. 
The lines show the $\Omega_{M}=0.5$ family of models from
Figure~\ref{focus}.  We have used constant values of \abeta=1.25 and 
log(O/H)=-3.9 for all galaxies at $z>0.1$ to compute their distance moduli. 
These values correspond to the mean values of the objects in our local sample 
that span ranges in $\lbeta$ and $\sigma$ covered by our intermediate
redshift sample and which are closest to those of the $z\sim3$ galaxies (cf. 
Section~5).  The large symbols show the average values for each sub-sample.  
The error bars show the mean error in distance modulus. Although our data-set 
cannot be used (nor is intended) to place significant constraints 
on the cosmological parameters, it is very helpful to
understand the limitations of the method. 

Probably the first thing one notices is the large scatter in the data.  
The rms dispersion in distance modulus for the local sample 
is $\rm \sigma[\Delta(m-M)]=0.52$ magnitudes. According
to MTM, typical errors for these galaxies are 
5\% in velocity dispersion and 
10\% in flux.  Adding errors of about 10\% in extinction and about 20\%
in abundance,  the expected scatter due just to observational errors
is 0.35~mag in distance modulus. Thus, there seems to be 
room for improvement and errors similar to those
of SNIa may be achievable with better quality data.


The second point is that the two high-redshift galaxies have distance moduli that 
are discrepant by more than one magnitude.  While the observational errors 
are indeed
large, this could also be due to our choice of extinction and metallicity.
These parameters enter with the  
same sign in Eq.1 so systematic changes of 0.2~dex in O/H and 0.2~mag in 
extinction (which correspond to $1\sigma$ deviations in the local sample) 
translate into shifts of 0.7~mag in distance modulus.  Notice that, since 
the maximum separation between $\Omega_{M}=0.2$ and $\Omega_{M}=1$ at z=3 
is about one magnitude (Figure~\ref{hubble}), it is crucial to have good 
measurements of extinction and abundance for these objects.

Finally we notice that even with our new zero-point calibration, 
the data for local HII galaxies are inconsistent
with the value of \hnot=65\hunits that results from SNIa.  
We believe that the discrepancy arises from systematic errors in the 
photometry of Giant HII regions which we are in the process of checking using
narrow-band CCD imaging.
Clearly, however, provided there are no systematic differences in the
photometric calibrations between local and distant objects, 
the determination of $\Omega$ is independent of \hnot.


We believe that using 
the new optical and IR spectrographs that are coming on-line on 8m-class 
telescopes it will be  possible to measure \fbeta\  to 10\% and $\sigma$ to
better than 5\% at z=3.  An accurate determination of $\Omega_{M}$,
with r.m.s. error about 0.05 seems therefore possible with samples of 
40-50 HII galaxies at z=1-3. The determination of extinction and metallicity 
at this redshift, however, will remain a challenging observational problem. 

\section[]{Conclusions}

Our exploration of the use of the magnitude-redshift method  
to determine the cosmological 
constant ($\Omega_{\Lambda}$)  and the mass density  of 
the Universe ($\Omega_M$) using HII galaxies led us to re-discover
the important focusing effect in $\Lambda$ for redshifts about 3.
This effect implies that the apparent magnitude of a standard candle
at redshifts z=2-3 has almost no dependence
on $\Omega_{\Lambda}$ for   $\Omega_M>0.2$.

Our strong conclusion is that using the redshift-magnitude diagram method
$\Omega_M$ can be measured independently of the value of
$\Omega_{\Lambda}$ by targeting the redshift range according to 
an estimate of the value of $\Omega_M$. In particular,
for small $\Omega_{M}$,
the optimum redshift is $z\sim 3$ where already a significant sample of
HII galaxies does exist (e.g.~Pettini \etal, 1998, Steidel \etal, 1998).

We also find that
the best range to determine  $\Omega_{\Lambda}$ using the redshift-magnitude
method is well away from the redshift region where the focusing
occurs, i.e. either $z < 1$ or $z >5$ .

Considering that we have very little control over the systematic
effects discussed above for galaxies at $z>0.1$, we find it quite
remarkable that the $\lbeta-\sigma$ relation established by MTM
for local HII galaxies is so well satisfied by objects of a wide range of
redshifts extending up to $z\simeq3$. Furthermore, the 
intermediate redshift sample itself shows a $\lbeta-\sigma$ correlation
similar to that found in the local Universe.  
Therefore, we are confident that HII galaxies
can potentially be used as cosmological probes out to redshifts z$=3 - 4$.

One should bear in mind that 
none of the intermediate and high redshift HII galaxies found
thus far has very strong emission lines. For most objects this
may be an effect of evolution plus the fact that we have not yet found the 
youngest galaxies at high redshifts.  This is not surprising because all
the intermediate and high redshift objects we have used in this paper
have been discovered using broad-band photometric techniques that
miss objects with very weak continua.  Searches are underway using
narrow band techniques that are revealing objects with redshifts
$z>3$ and strong Lyman-$\alpha$ lines (Hu \etal,  1998).  We think
that many of these may in fact be young HII galaxies. Using the high 
efficiency IR spectrographs that are becoming available in the new 
generation of 8-10m telescopes it will be possible to determine the \hb\ 
line widths, luminosities, and equivalent widths of these objects over a wide 
range of luminosities with high accuracy.  This will allow for the
first time using the distance estimator to probe the cosmological
parameters out to unprecedented distances.

\section*{Acknowledgments}

We enjoyed fruitful discussions with George Efstathiou, Max Pettini, Rafael 
Guzm\'an and Dave Koo. We are 
grateful to Benoit Joguet who helped with the reduction of the 
low metallicity intermediate z HII galaxies data, 
and Eduardo Telles for providing us with his fundamental plane diagram and
allowing us to include it in this paper. It is a pleasure to acknowledge 
the support and hospitality of the Guillermo Haro programme
for Advanced Astrophysics at INAOE, where this paper was written.
We thank our anonymous referee for a number of valuable comments and 
criticisms that have not only improved the paper, but also helped us to 
improve our understanding of the observational challenges that lie
ahead of us.

\bsp

\end{document}